\documentclass[useAMS,usenatbib,usegraphicx]{mn2e}

\def\kms{km s$^{-1}$} \def\msun{M$_{\sun}$}  \def\aap{A\&A}
\def\apjl{ApJ} \def\apj{ApJ} \def\apjs{ApJS} \def\aj{AJ} \def\mnras{MNRAS}
\def\nat{Nature} \def\pasp{PASP} 

\title[The Runaway Binary LP 400-22]{The Runaway Binary LP 400-22 is Leaving the Galaxy}

\author[M. Kilic et al.]{
Mukremin Kilic$^1$\thanks{kilic@ou.edu},
A. Gianninas$^1$,
Warren R. Brown$^2$,
Hugh C. Harris$^3$, 
Conard C. Dahn$^3$,
\newauthor M. A. Ag\"{u}eros$^4$, 
Craig O. Heinke$^5$,
S. J. Kenyon$^2$,
J. A. Panei$^6$,
Fernando Camilo$^{4,7}$ \\
$^1$Homer L. Dodge Department of Physics and Astronomy, University of Oklahoma, 440 W. Brooks St., Norman, OK, 73019, USA \\ 
$^2$Smithsonian Astrophysical Observatory, 60 Garden St, Cambridge, MA 02138, USA\\ 
$^3$US Naval Observatory, Flagstaff Station, 10391 West Naval Observatory Road, Flagstaff, AZ 86001, USA\\
$^4$Columbia University, Department of Astronomy, 550 West 120th Street, New York, NY 10027, USA\\ 
$^5$Department of Physics, CCIS 4-183, University of Alberta, Edmonton, AB, T6G 2E1, Canada \\
$^6$Instituto de Astrofisica La Plata, IALP, CONICET-UNLP, Argentina \\
$^7$Arecibo Observatory, HC3 Box 53995, Arecibo, PR 00612, USA 
}

\begin{document}

\maketitle

\begin{abstract}

We present optical spectroscopy, astrometry, radio, and X-ray observations of
the runaway binary LP 400-22. We refine the orbital parameters
of the system based on our new radial velocity observations. Our parallax data
indicate that LP 400-22 is significantly more distant (3$\sigma$ lower limit
of 840 pc) than initially predicted. LP 400-22 has a tangential velocity in
excess of 830 \kms; it is unbound to the Galaxy. Our radio and X-ray
observations fail to detect a recycled millisecond pulsar companion, indicating
that LP 400-22 is a double white dwarf (WD) system. This essentially rules out a supernova
runaway ejection mechanism. Based on its orbit, a Galactic center origin is also
unlikely. However, its orbit intersects the locations of
several globular clusters; dynamical interactions between LP 400-22 and other
binary stars or a central black hole in a dense cluster could explain the
origin of this unusual binary.

\end{abstract}


\section{INTRODUCTION}

Dynamical processes involving n-body interactions in dense star clusters,
supernova explosions in tight binary systems, or interactions with the
black hole at the Galactic center can eject stars from their birthplace at high
velocities. In the first scenario, most of the runaway stars ejected from
clusters through three-, four-, or n-body interactions \citep{poveda67} are
expected to be single stars. However, a small fraction should be tight binaries
ejected through close encounters \citep{gies86}. Binary-binary encounters may
disrupt one or both systems, but in 10\% of the cases, two binaries are ejected
\citep{mikkola83}.  In the supernova ejection scenario \citep{blaauw61}, all
runaway binary systems contain the remnant of a supernova explosion, a neutron
star or a black hole.  However, in the majority of the cases, these companions
cannot be detected due to selection effects, e.g. small radial velocity
variations due to the low-mass of a neutron star compared to the runaway OB
stars and the short lifetimes of radio pulsar companions \citep{portegies00}.

Hypervelocity stars \citep[HVSs,][]{brown05,brown07,edelmann05} with velocities
even higher than the runaway stars, are likely ejected from the Galactic center
\citep{hills88, brown12}.  So far, there are no binary HVSs known. However,
several HVSs have main-sequence lifetimes significantly shorter than their
travel time from the Galactic center. \citet{perets09} suggests that such stars
may be the result of a merger of a HVS binary system that rejuvenates itself
after its ejection from the Galactic center. HE 0437$-$5439
\citep{edelmann05,brown10} and US 708 \citep{hirsch05} are two such systems,
where the progenitor HVS binaries might have been ejected due to triple
disruptions and other dynamical interactions with stars or black holes.

\citet{lu07} and \citet{sesana09} propose that the discovery of HVS binary stars
would indicate the existence of a binary black hole at the Galactic
center.  In their model, HVS binaries with velocities $\sim$ 1,000 \kms\ can be
ejected from the Galactic center due to interactions with a binary black
hole. Even though tidal disruption of a hierarchial triple star system by a
single black hole may lead to a HVS binary, they find the ejection rate
from this mechanism is negligible \citep[although see the discussion
in][]{perets09}.

In this paper we revisit the unusual runaway binary LP 400-22 (WD 2234+222
at $l=86.5^{\circ}$ and $b=-30.7^{\circ}$). \citet{kawka06}
identified LP 400-22 as a fast moving, extremely low-mass (ELM, 0.17 $M_{\odot}$) WD at a
distance of 430 $\pm$ 45 pc.  \citet{kilic09} and \citet{vennes09} obtained
follow-up radial velocity observations, which demonstrated that LP 400-22 is a
double degenerate binary system with an orbital period of $\approx$ 1 day. Based
on the available astrometric data, \citet{kilic09}
estimated that the probability of a Galactic center origin is 0.1\% and
concluded that LP 400-22 is most likely a halo star with an unusual orbit. Here
we present additional radial velocity, astrometry, X-ray, and radio observations
of LP 400-22 and revisit its origin. Our observations are discussed in Section
2, whereas the revised binary parameters and its Galactic orbit are discussed in
Sections 3 and 4, respectively.

\section{OBSERVATIONS}

\subsection{Optical Spectroscopy}

We used the 6.5m MMT equipped with the Blue Channel Spectrograph to obtain
additional optical spectroscopy of LP 400-22 over several nights in September
2009 and November 2010. Our observing
and reduction procedures are similar to those of \citet{kilic09}. The only
difference is that most of the new spectra were obtained using a 1.25$\arcsec$
slit, instead of the 1$\arcsec$ slit used in our earlier work.

\begin{figure}
\vspace*{-1.7in}
\hspace{-0.7in}
\includegraphics[width=4.5in,angle=0]{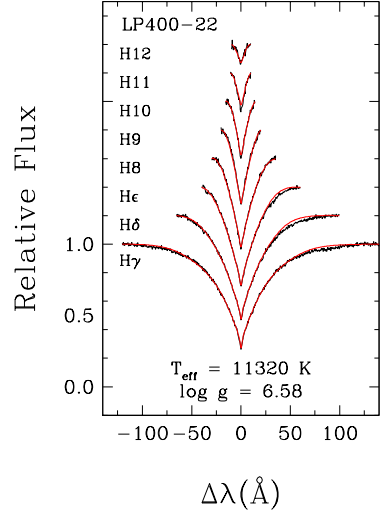}
\vspace*{-1.7in}
\caption{Spectral fits (red lines) to the MMT spectrum (black lines) of LP 400-22.
\label{fig:spec}}
\end{figure}

In addition to measuring radial velocities, we use our MMT data to revise the
atmospheric parameters of the visible ELM WD in the system. Figure \ref{fig:spec} shows
the fits to the composite spectrum using the pure hydrogen atmosphere models from \citet{tremblay09}.
The best fit temperature and gravity with the formal errors are $T_{\rm eff} = 11320 \pm 40$ K
and $\log{g} = 6.58 \pm 0.01$. The best-fit spectroscopic distance estimate is 350 pc.
The temperature estimate is consistent with the previous analysis \citep{kawka06,kilic09,vennes09}.
However, the gravity estimate is 0.16-0.28 dex higher. The improved Stark broadening profiles of
the \citet{tremblay09} models lead to an increase of $\sim$0.1 dex in surface gravity for average
mass WDs. We perform a similar analysis of the LP 400-22 spectrum using both old and the new models and
find that the new models
increase the best-fit $\log{g}$ estimate by 0.14 dex. Hence, the difference between the previous
surface gravity measurements and ours is mostly due to the improved Stark broadening profiles in the new models.  

\subsection{Astrometry}

\begin{figure*}
\includegraphics[width=5.0in,angle=0]{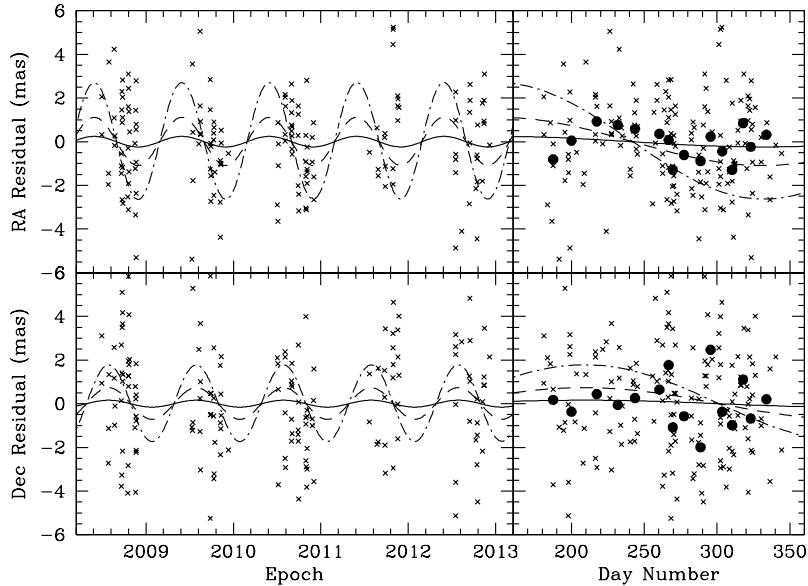}
\vspace{-0.5in}
\caption{The observed parallactic motion in RA (top panel)
and Dec (bottom panel), after fitting the astrometry
for parallax and proper motion and subtracting
the proper motion. The data have the correction
to absolute parallax of 0.68 mas added to the observed
residuals, as if the reference stars were at infinite
distance. The solid curve show the best fit parallax,
the dashed fit shows the 3-sigma upper limit to the
parallax, and the dot-dash curve shows the parallax
for the spectroscopic distance of 347 pc from Fig. \ref{fig:spec}.
Right panels show the same parallax data folded on a 1 year period,
with 10-point-means added (filled circles).
\label{fig:pi}}
\end{figure*}

Astrometric observations of LP 400-22 spanning five seasons
were obtained as part of the USNO parallax program. Data acquisition
and reduction procedures are described in \citet{dahn02}. Figure
\ref{fig:pi} presents the observed parallactic motion for LP 400-22.
Using 159 frames taken on 92 nights and 20 reference stars, we measure
a relative proper motion of $\mu = 208.5 \pm 0.2$ mas yr$^{-1}$ at position
angle $74.07^{\circ} \pm 0.15^{\circ}$, and a relative parallax of $-0.42 \pm
0.30$ mas. Using the colors and magnitudes of the reference-frame stars, we
estimate the absolute parallax to be 0.26 $\pm$ 0.31 mas.
Given the relatively small parallax and its large error, we cannot constrain
the actual distance to LP 400-22, but we can still use the 3$\sigma$ lower limit of
our parallax measurement to constrain the physical properties of the system.
The nominal distance of 3.8 kpc would correspond to a tangential velocity of 3800 \kms,
which is extremely unlikely.
The 2$\sigma$ and 3$\sigma$ lower limits on the distance are 1130 and 840 pc, respectively. 
Our proper motion and distance measurements imply a 3$\sigma$ lower limit on tangential
velocity of 830 \kms, higher than the escape velocity from the Galaxy.

\subsection{Radio}

We observed LP 400-22 on 2008 Mar 28 for 22.5 min with the Berkeley-Caltech
Pulsar Machine \citep[BCPM][]{backer97} on the Green Bank Telescope. At the
central observing frequency of 820 MHz, the BCPM provided 48 MHz of bandwidth
split into 96 spectral channels; for each channel we recorded total power samples
every 72 $\mu$s. The data reduction was similar to that described in \citet{agueros09b}:
we used the standard search techniques implemented in the PRESTO software package
\citep{ransom01}. The dispersion measure range for the search was $0-115$ cm$^{-3}$ pc.

We note that while orbital motions can affect the apparent spin period of a pulsar (i.e.,
if the integration time is a significant fraction of the binary orbital period), here
the integration time is only about 1\% of the orbital period, so that the assumption
of constant apparent acceleration built into PRESTO should hold. No convincing pulsar
signal is detected in our data.

\subsection{X-ray}

We obtained observations of LP 400-22 with XMM (ObsID 0553440201, on 2008 Nov 20)
and Chandra (ObsID 9962, on 2009 July 28).  We analyzed the pre-processed
(PPS or event2) datasets using SAS v. 11.0 and CIAO v. 4.3.  We checked all data
for background flares; finding none, we used the whole datasets (livetimes of
1028 s, 4438 s, 6520 s, and 6520 s for Chandra ACIS, XMM pn, MOS1, and MOS2
respectively).  We filtered the XMM datasets following standard
procedures\footnote{http://heasarc.gsfc.nasa.gov/docs/xmm/abc/}; PATTERN $\leq$12
and standard FLAG values for the MOS and pn detectors.  We created images for
each detector in the energy range of 0.5 to 2 keV.

We calculate the minimum flux expected, should the companion to LP 400-22 be a
(recycled) neutron star, with an X-ray luminosity ($L_X$(0.3-2
keV)$=1.5\times10^{30}$ ergs cm$^{-2}$ s$^{-1}$) and spectrum (134 eV blackbody)
equal to that of the faintest millisecond pulsar in 47 Tuc
(\citealt{heinke05,bogdanov06}; see \citealt{agueros09a} for the rationale).
Assuming a distance of 1.5 kpc and $N_H=4.5\times10^{20}$ cm$^{-2}$, we predict 0.75, 10, 6,
and 6 counts in each detector, using the simulator
PIMMS\footnote{http://asc.harvard.edu/toolkit/pimms.jsp} and the exposure times
and energy range above.

For the Chandra observation, we searched for emission from a 2$\arcsec$ circle centered
on LP 400-22's position in the 0.5-2 keV band, but found zero counts. We
searched for emission from 10$\arcsec$ circles in the XMM images, and found
one event in the pn image, and one event in the combined MOS1 and MOS2 image.
From XMM encircled energy calculations,
we find that 60\% of the encircled energy should be located within these 10$\arcsec$
extraction regions, so we revise our predicted counts to 6 counts in the
searched region of the pn image, and 3.6 counts in this region of the combined
MOS image.  Combining the XMM and Chandra studied regions, we expect to see 10.4
counts, and only see two.  Using Poisson statistics \citep{gehrels86}, we
conclude that we can rule out the hypothesis of a minimal-flux millisecond
pulsar at $>$99.5\% confidence, even at the extreme distance of 1.5 kpc.
We can rule-out millisecond pulsar companions at 90\% confidence level out to about 2 kpc.
Clearly, LP 400-22 must be within 2 kpc of the Sun for it to have a reasonable tangential
velocity of $\leq2,000$ \kms. Therefore, we conclude that the companion must be a WD.

\section{RESULTS}

\subsection{Orbital Parameters}

\begin{figure}
\includegraphics[width=3.3in,angle=0]{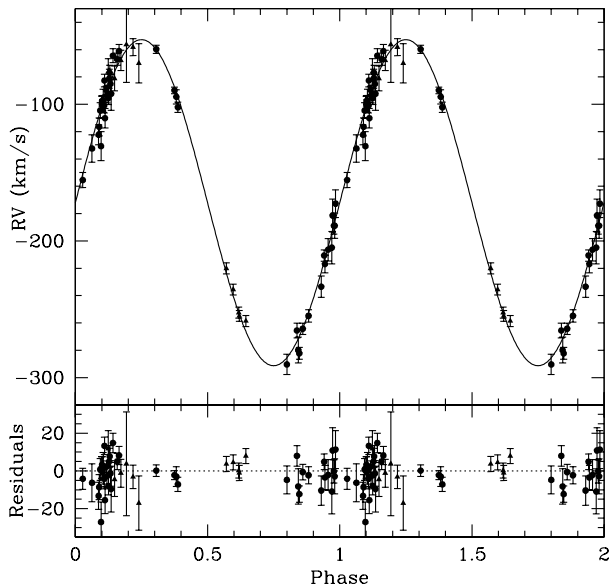}
\caption{Radial velocity measurements of LP 400-22 from \citet{kilic09},
\citet{vennes09}, and this work. The solid line represents the best-fit
model for a circular orbit with a period of 1.01014 d and $K=119.3$ \kms.
\label{fig:radvel}}
\end{figure}

We provide improved orbital parameters of the LP 400-22 binary using the radial
velocity measurements from \citet{kilic09}, \citet{vennes09}, and our new
spectroscopy data. Table 1 lists our 29 new radial velocity measurements.
Figure \ref{fig:radvel} presents 53 radial velocity measurements obtained
over 1500 days and the best-fit orbit with $P = $1.01014(5)
d, $K = $119.3(8) \kms, and systemic velocity
$\gamma = -$172.0(5) \kms. The revised mass function is
$f = $0.1778(36). These orbital parameters are consistent with $P = $1.01016(5) d
and $f = $0.180(9) from \citet{vennes09} within the errors. 

\begin{table}
\centering
\caption{New Radial Velocity Measurements for LP 400-22}
\begin{tabular}{rr}
\hline
HJD$-$2455000 & $v_{helio}$ \\
(days) & (\kms) \\
\hline
 94.81508 & $-$102.2 $\pm$  3.8 \\
100.61535 &  $-$77.1 $\pm$  5.7 \\
100.64317 &  $-$67.0 $\pm$  4.4 \\
100.65032 &  $-$61.1 $\pm$  4.8 \\
100.79241 &  $-$59.8 $\pm$  3.0 \\
101.59483 & $-$100.0 $\pm$  5.7 \\
101.59985 &  $-$97.4 $\pm$  8.4 \\
101.60701 & $-$110.2 $\pm$  7.1 \\
101.61413 &  $-$93.5 $\pm$  5.5 \\
101.62249 &  $-$81.2 $\pm$  6.9 \\
101.62961 &  $-$92.3 $\pm$ 12.1 \\
101.63677 &  $-$64.5 $\pm$  5.1 \\
101.87234 &  $-$89.8 $\pm$  2.4 \\
102.59303 & $-$122.3 $\pm$  7.2 \\
102.59914 & $-$104.5 $\pm$  5.8 \\
102.60759 & $-$100.5 $\pm$  3.9 \\
102.61494 &  $-$82.6 $\pm$  4.6 \\
102.62206 &  $-$89.7 $\pm$  3.1 \\
102.63034 &  $-$95.4 $\pm$  4.7 \\
102.88991 &  $-$94.6 $\pm$  5.2 \\
511.56650 & $-$206.4 $\pm$  8.2 \\
511.58061 & $-$204.9 $\pm$ 11.9 \\
511.58308 & $-$181.4 $\pm$ 12.1 \\
511.59159 & $-$188.9 $\pm$  9.3 \\
511.59425 & $-$172.7 $\pm$ 10.1 \\
512.68415 & $-$132.4 $\pm$ 10.0 \\
512.71175 & $-$116.5 $\pm$  6.6 \\
512.75448 &  $-$85.1 $\pm$  6.8 \\
513.72902 & $-$130.7 $\pm$ 10.6 \\
\hline
\end{tabular}
\end{table}

\subsection{Physical Properties}

\begin{figure*}
\includegraphics[width=3.5in,angle=-90]{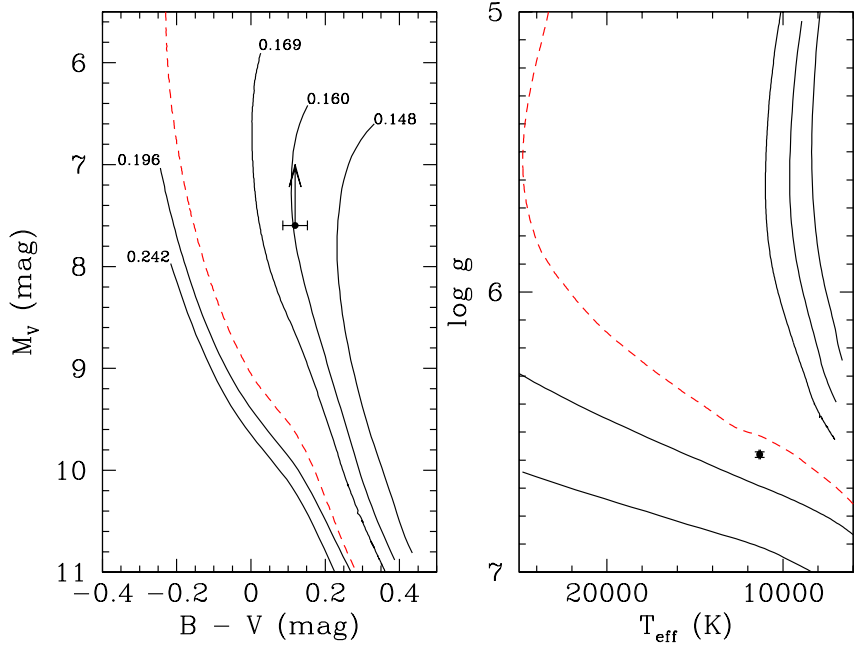}
\caption{LP 400-22 on a color-magnitude diagram (left) and a $T_{\rm eff}$ vs. $\log {g}$
diagram. Evolutionary model sequences for 0.15-0.24 $M_{\odot}$ He-core WDs
\citep[solid lines,][]{serenelli01} and 0.19 $M_{\odot}$ WDs
\citep[dashed lines,][]{panei07} are also shown.
\label{fig:models}}
\end{figure*}

Figure \ref{fig:models} compares the luminosity, effective temperature, and
surface gravity of the primary star in LP 400-22 to the evolutionary
sequences for 0.15-0.24 $M_{\odot}$ WDs from \citet{serenelli01} and
\citet{panei07}. The $3\sigma$ lower limit on distance indicates
an absolute magnitude of $M_{\rm V}=7.6$ mag, $\log {L/L_{\odot}} = -1.05$,
and $R = 0.099 R_{\odot}$. These are consistent with the \citet{serenelli01}
models for a 0.16 \msun\ WD with a thick hydrogen envelope. However,
these models predict
a surface temperature of 9600 K and $\log {g} = 5.6$, an order of magnitude
lower in surface gravity than the best-fit model shown in Figure
\ref{fig:spec}.

Clearly, there is no single model that matches all of the properties of
the primary star in LP 400-22. A 0.19 $M_{\odot}$ WD model can match the
the best-fit temperature and surface gravity measurements, but it
underpredicts the absolute magnitude, radius, and distance.
Such a discrepancy between the model and parallax distance measurement for an average mass
(0.6 \msun) DA WD could indicate the existence of spectroscopically invisible helium.

Increased pressure broadening of hydrogen lines due to the presence of helium can imitate
a higher surface gravity.
GD 362 is perhaps the best example of such a system, where the initial model atmosphere
analysis showed that the optical spectrum can be explained by a massive
(1.2 \msun) DA WD \citep{gianninas04}. However, follow-up high resolution observations
\citep{zuckerman07} and a parallax measurement \citep{kilic08} showed that GD 362
indeed has a helium dominated atmosphere and that it is not a massive WD. Since He becomes
transparent below about 11,000 K, it is possible to hide significant amounts of He in
the atmosphere of LP 400-22.
To investigate this possibility and to check whether a mixed H/He atmosphere solution would
give results consistent with the parallax measurement, we computed model atmospheres with
He/H ratios ranging from 0 to 100. 

\begin{figure}
\vspace{-0.9in}
\hspace{-1.2in}
\includegraphics[width=4.3in,angle=-90]{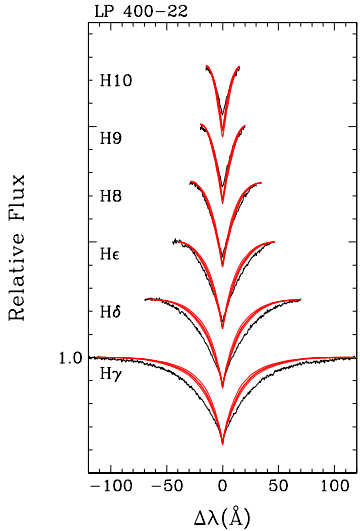}
\vspace{-0.7in}
\caption{The MMT spectrum (black lines) of LP 400-22 compared to mixed
atmosphere models with $T_{\rm eff}=9600$ K, $\log{g}=5.6$, and
He/H = 0, 1, 10, and 20 (from top to bottom, red lines). The Balmer
line profiles for these mixed atmosphere models are essentially identical to
each other due to the low surface gravity.\label{fig:linecomp}}
\end{figure}

Figure \ref{fig:linecomp} shows the Balmer line profiles of LP 400-22 compared to mixed atmosphere models with $T_{\rm eff}=9600$ K, $\log{g}=5.6$ (see the above discussion), and He/H = 0, 1, 10, and 20.
None of these models provide a reasonable fit to the observed
spectrum of LP 400-22. This is because the addition of helium to a $\log{g}=6$ atmosphere
does not have the same effect as that of a $\log{g}=8$ or 9 atmosphere. 
In addition, the spectral energy distribution analysis of \citet{kawka06} using ultraviolet
and optical photometry is in excellent agreement with the Balmer line analysis using pure H models.
Hence, a temperature as low as 9600 K can be ruled out from both spectroscopy and photometry.

\begin{figure*}
\vspace{-0.7in}
\includegraphics[width=5.0in,angle=-90]{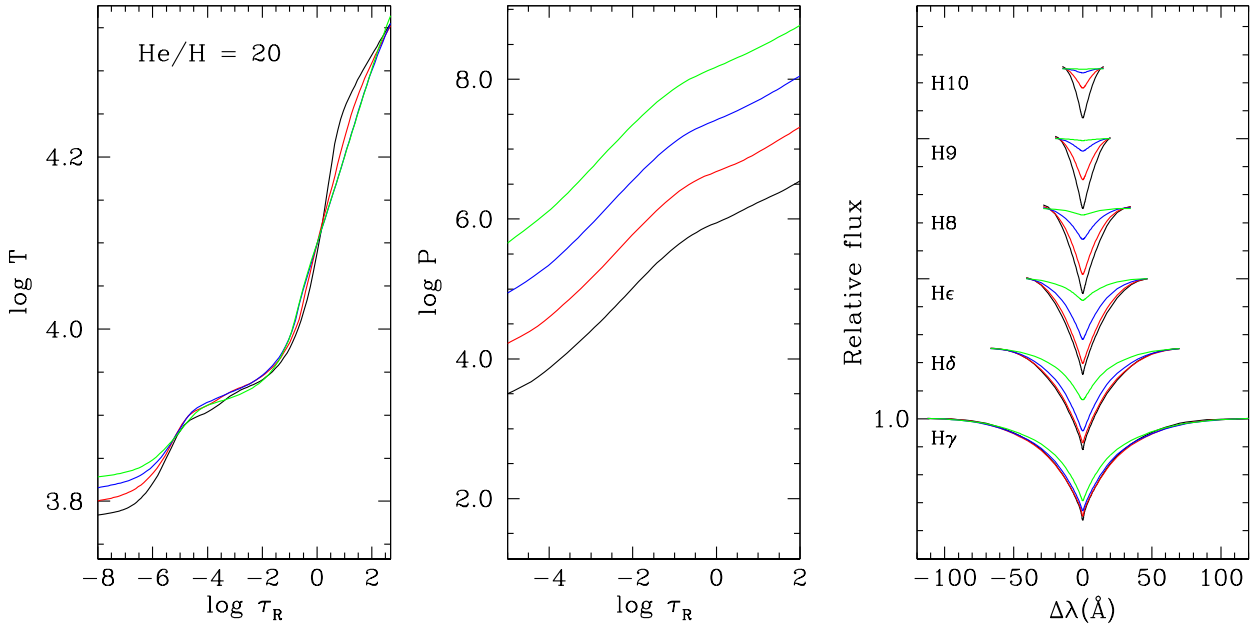}
\vspace{-0.8in}
\caption{Temperature (left), pressure (middle), and Balmer line profiles (right)
of model atmospheres with He/H = 20, $T_{\rm eff}=11,500$ K, and $\log{g}=$ 6, 7, 8, and 9
(from bottom to top).
\label{fig:pressure}}
\end{figure*}

Figure \ref{fig:pressure} shows the temperature, pressure,  and Balmer line profiles
of mixed H/He atmosphere models with He/H = 20 and
$\log{g}=6,7,8,$ and 9. The pressure for the $\log{g}=6$ model is two orders of magnitude
smaller than it is for the $\log{g}=9$ model. The results are similar for models with different
He/H ratios, e.g. He/H = 50, 100. Hence, the effect of helium on the atmospheric
structure and the Balmer line profiles is negligible for low surface gravity models
appropriate for ELM WDs, including LP 400-22.  
Figures \ref{fig:linecomp} and \ref{fig:pressure} demonstrate that the addition of He in our spectral models for LP 400-22 does not
resolve the discrepancy between the observed properties and the evolutionary models.

An alternative explanation for the discrepancy in the spectroscopic distance estimate is an
unresolved DA + DC WD system, where the DC WD companion is bright enough to contribute to the
spectral continuum but not to the Balmer line profiles. We tested this scenario by fitting the observed
spectrum as a composite DA + DC system. We fixed the $\log{g}$ of the DC component to
$\log{g}=7.5$ or 7.7 (0.4 or 0.5 $M_{\odot}$) and allowed the other three parameters, temperature
and surface gravity of the DA component and the temperature of the DC component, to vary.
Our fitting algorithm achieves a good fit to the observed spectrum only if the DC
component is cool enough ($T_{\rm eff}<3500$ K) that it essentially does not contribute any
light to the system. We do not find any plausible solutions involving a composite DA + DC spectrum.

\citet{althaus01} and \citet{serenelli01} predict that $M \geq 0.2 M_{\odot}$ WDs suffer
from diffusion-induced hydrogen shell flashes that lead to thin hydrogen
envelopes. The surface gravity and temperature measurements for LP 400-22 place it
right in the transition region between the objects with and without hydrogen shell flashes
(see Fig. 3). \citet{panei07} argue that the lower mass limit for
shell flashes is 0.17 $M_{\odot}$. Below this mass limit, WDs have
a thick hydrogen surface layer that provides energy through stable hydrogen
burning. This extra energy source slows down the evolution of the lowest
mass WDs, keeping them luminous for longer than expected. 
Based on the \citet{panei07} models, the temperature and surface gravity
measurements for LP 400-22 imply a $\approx0.19 M_{\odot}$ WD with a thin
H envelope of $\log{M/M_{\star}}=-3.1$. On the other hand, 
its luminosity and colors do not match any of the \citet{panei07} cooling
tracks; LP 400-22 may be a pre-WD.

The mass estimate for LP 400-22 ranges from 0.16 $M_{\odot}$
to 0.19 $M_{\odot}$.
The observed parallax implies that LP 400-22 is significantly more luminous
than initially predicted from the models. However, given the uncertainties in models and
specifically the transition from hydrogen shell flashes to stably burning envelopes for
$\approx0.2$ \msun\ WDs, it is perhaps not surprising to find discrepancies between the model
expectations and the observed properties. Regardless of these issues,
the observed surface gravity of the
primary star in LP 400-22 shows that it is clearly a WD, and its brightness suggests that
it is likely to have a thick envelope with stable hydrogen burning. 

We note that the ELM WD in the eclipsing double WD binary NLTT 11748 \citep{steinfadt10}
looks like a cooler version of the primary star in LP 400-22. NLTT 11748 has essentially the
same surface gravity as LP 400-22, but the primary WD has $T_{\rm eff}=8690 \pm 140$ K.
NLTT 11748 was also observed as part of the USNO Parallax Program, and the distance
measurement agrees with model expectations for that star (H. C. Harris 2013,
private communication). Hence, the difference between the NLTT 11748 and LP 400-22 ELM WDs
may be a slight difference in mass and a significant difference in the thickness of the surface hydrogen layer.   

\subsection{The Galactic Orbit}

The lack of a signature of a millisecond pulsar companion in the radio and
X-ray observations demonstrate that LP 400-22 is a double WD binary system
containing an ELM WD and an invisible higher mass ($M\geq0.39$\msun) WD
companion at an inclination larger than 33$^\circ$. 

\begin{figure}
\includegraphics[width=3.4in,angle=0]{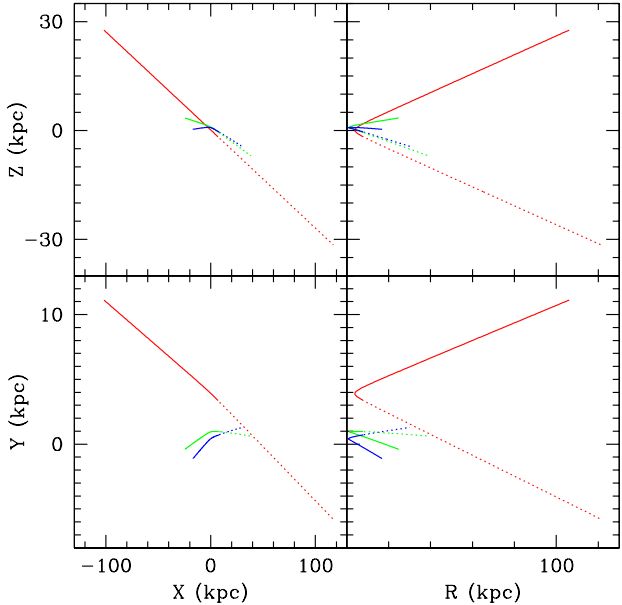}
\caption{Galactic orbit of LP400-22 for a distance of 3800 pc (red),
1130 pc (green, $2\sigma$ lower limit), and 840 pc (blue, $3\sigma$ lower limit).
The solid and dotted lines show the past and future orbits for 30 Myr.}
\label{fig:allorbits}
\end{figure}

Figure \ref{fig:allorbits} presents its Galactic orbit assuming a static disk-halo-bulge
potential \citep{kenyon08}, 250 \kms\ solar rotation \citep{reid09}, the local
standard of rest as defined by \citet{schonrich10}, and $X=8$ kpc and $Z=20$ pc
for the location of the Sun \citep{reed06}. Even at the 3$\sigma$ lower limit
of 840 pc, LP 400-22 is unbound.

For the three trajectories shown in Figure \ref{fig:allorbits}, the closest
pericenter passage occurs 2-9 Myr ago at distances of 0.9-3.9 kpc.
Hence, a Galactic center origin is effectively ruled out for LP 400-22. The main
problem is that the $W$ component of the velocity, $-140$ ($-220$) \kms\ for the
$3\sigma$ ($2\sigma$) lower limit on distance, is too large for a trajectory
originating in the Galactic center. 
We note that the use of a different Galactic potential, i.e. that of \citet{gnedin05},
does not change any of our conclusions.

\section{DISCUSSION}

LP 400-22 is an unbound runaway binary WD system. There are other binary
runaway systems known; roughly 5\% to 26\% of runaway O-type stars are binary
systems \citep{mason98}. However, LP 400-22 is unique, since it is the only runaway
binary WD system currently known and it has an extremely large space velocity
compared to typical runaway stars. 

Unlike the neutron star
companions to OB stars, a neutron star companion to LP 400-22 should be
detected as a recycled millisecond pulsar through radial velocity,
radio, and X-ray observations. Our observations indicate that LP 400-22 binary
has a circular orbit with no evidence of a neutron star companion. In addition,
$<$1\% of runaways are expected to receive velocity kicks in excess of 200 \kms
\citep{portegies00}. The observed space velocity of LP 400-22 and the non-detection
of a neutron star companion indicate that the supernova ejection mechanism cannot
explain the origin of the LP 400-22 binary.

Hypervelocity binary systems may form due to the tidal disruption of
a hierarchical triple star by the central black hole or the interaction
of a binary star with a binary black hole at the Galactic center.
\citet{lu07} estimate that the latter formation channel can form hypervelocity
binary systems with velocities up to 1,000 \kms. However, LP 400-22 has a Galactic
orbit that is inconsistent with a Galactic center origin. Hence, the Galactic
center ejection mechanism through interactions with the central black hole(s)
also cannot explain the origin of the LP 400-22 binary. 

An alternative birthplace where interactions with an intermediate mass black hole can take
place and the number density of the stellar population is high enough for frequent
close encounters is globular clusters. Multi-body interactions in a dense star
cluster can in principle explain the origin of LP 400-22. Some runaway stars can
be traced back to
their birthplace because of their short main-sequence lifetimes and their proximity
to known clusters, associations, or H II regions. In the case of LP 400-22, its WD age
and total age are uncertain, and there is no way to confirm its birthplace given
the uncertainties in its distance measurement. Nevertheless, an origin in a
globular cluster is possible.

\begin{figure}
\includegraphics[width=3.4in,angle=0]{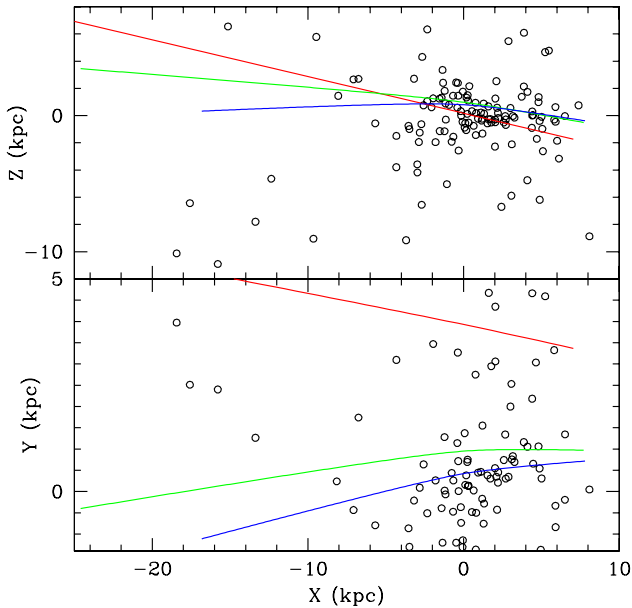}
\caption{Galactic orbit of LP400-22 for the past 30 Myr.
The symbols are the same as in Figure \ref{fig:allorbits}.
Open circles show the positions of the known globular clusters.}
\label{fig:orbit}
\end{figure}

Figure \ref{fig:orbit} shows the Galactic orbit for LP 400-22 for the past 30 Myr along
with the positions of the known globular clusters \citep{harris96}. For a distance of
840 pc, 4.2 Myr ago LP 400-22 was within 170 pc of the globular cluster 2MASS-GC01.
Given the uncertainties in distance, it is not possible
to locate the cluster that LP 400-22 definitely came from. However, this figure
demonstrates that there are several globular clusters in the vicinity of LP 400-22's
trajectory that may explain its origin.

There are only a handful of hypervelocity stars with accurate proper motion measurements.
The problem is that many of them are located at $\sim$ 100 kpc; space based
telescopes are required to measure their proper motion. \citet{brown10} used the {\em Hubble
Space Telescope} to measure the proper motion of the hypervelocity star HE 0437-5439, and
demonstrated that its velocity vector points directly away from the Galactic center.
Similar observations of the other known hypervelocity stars are required to directly
link them to the Galactic center. 

There are two cases where accurate proper motions are available and they are
inconsistent with a Galactic center origin. HD 271791 is an 11 $M_{\odot}$ B-type star
with a velocity larger than the Galactic escape velocity. \citet{heber08} show that
its kinematic properties rule out a Galactic center origin, and they are more
consistent with the formation in the outer disk. Similarly, SDSS J013655.91+242546.0
is a 2.5 $M_{\odot}$ A-type main-sequence star, likely unbound to the Galaxy
and with an origin near the outer Galactic disk \citep{tillich09}. LP 400-22
joins this group of runaway stars for which a Galactic center origin is ruled out based
on proper motion measurements. 

\citet{gualandris07} and \citet{gvaramadze08} argue that close encounters between
two hard binaries or interactions with an intermediate-mass ($\sim10^3 M_{\odot}$)
black hole in a dense environment can explain hypervelocity stars like HE 0437-5439,
assuming that it originated in a star cluster in the Large Magellanic Cloud (LMC).
Even though we now know that HE 0437-5439 did not originate in the LMC \citep{brown10},
the same arguments can be used to explain the origin of the LP 400-22 binary system in a
Milky Way globular cluster. Hard binary interactions can eject single stars with
velocities up to 1,400 \kms \citep{leonard91}. The ejection velocity can be as large
as 2,300 \kms, if the binaries involved in the interaction consist of low-mass systems
that have already gone through common-envelope evolution \citep{gvaramadze08}. Even though
detailed dynamical simulations for the LP 400-22 binary are currently not available,
it is likely that LP 400-22 formed through multi-body interactions involving hard
binaries and/or an intermeidate-mass black hole in a dense cluster.

\section{CONCLUSIONS}

We refine the orbital parameters of the intriguing runaway binary WD LP 400-22. Based on
our parallax measurements, we demonstrate that this binary is unbound to
the Galaxy. We estimate its Galactic orbit using a static disk-halo-bulge potential and find
a Galactic center origin unlikely. No neutron star companion is detected in the radio and
X-ray data, indicating that a supernova ejection mechanism is ruled out for this system.
The only remaining explanation for the unusual space velocity of LP 400-22 is
multi-body interactions involving hard binary systems or the disruption of a hierchical
triple system by an intermediate-mass black hole in a dense cluster. LP 400-22's
trajectory intersects several known globular clusters. However, due to the relatively
uncertain distance measurement, we cannot link LP 400-22 with any single globular cluster.
Further follow-up parallax observations at the USNO and with the GAIA satellite will be extremely
helpful in constraining the Galactic orbit for LP 400-22. In addition, dynamical simulations
specifically designed to study the LP 400-22 binary WD system will be useful for
understanding its origin in a dense cluster and its implications for the existence
of intermediate-mass black holes in globular clusters.

\section*{Acknowledgements}
We thank P. Bergeron for useful discussions on mixed model atmospheres
and S. Vennes for a constructive referee report.
MK is thankful to the University of Oklahoma Research Council and the College
of Arts and Sciences for a Junior Faculty
Fellowship. CH is supported by an NSERC Discovery Grant, and by an Alberta Ingenuity
New Faculty Award. Part of the observations reported here were obtained at the
MMT Observatory, a joint facility of the Smithsonian Institution and the University of Arizona.
The Robert C.\ Byrd Green Bank Telescope is operated by the
National Radio Astronomy Observatory, which is a facility of the US National Science
Foundation operated under cooperative agreement by Associated Universities, Inc.

{\it Facilities: MMT, CXO, XMM, GBT}

\end{document}